# Exact microscopic theory of electromagnetic heat transfer between a dielectric sphere and plate

Clayton Otey[1] and Shanhui Fan[2]

[1]Department of Applied Physics and [2]Department of Electrical Engineering, Stanford University, Stanford, California, 94305 U.S.A.

**Near-field electromagnetic heat transfer holds great potential for the advancement of nanotechnology. Whereas far-field electromagnetic heat transfer is constrained by Planck's blackbody limit, the increased density of states in the near-field enhances heat transfer rates by orders of magnitude relative to the conventional far-field limit [1-3]. Such enhancement opens new possibilities in numerous applications [4], including thermal-photo-voltaics [5], nano-patterning [6], and imaging [7]. The advancement in this area, however, has been hampered by the lack of rigorous theoretical treatment, especially for geometries that are of direct experimental relevance. Here we introduce an efficient computational strategy, and present the first rigorous calculation of electromagnetic heat transfer in a sphere-plate geometry, the only geometry where a transfer rate beyond the blackbody limit has been quantitatively probed at room temperature [8-10]. Our approach results in a definitive picture unifying various approximations previously used to treat this problem, and provides new physical insights for designing experiments aiming to explore enhanced thermal transfer.**

In order to advance the understanding of near-field electromagnetic heat transfer, two groups have recently measured the electromagnetic heat flux between a micron scale sphere and a plate with sub-micron vacuum separations between the surfaces. In accordance with qualitative theoretical predictions, they observed substantial enhancement beyond the blackbody limit [8-10]. On the theoretical side, it has been well established that near-field electromagnetic heat transfer can be rigorously modeled using classical macroscopic fluctuational electrodynamics (CMFED) [11]. However, exact CMFED results have been largely restricted to planar geometries [1]. The two sphere heat transfer problem was tackled only recently [12], and the experimentally important sphere-plate geometry has been elusive; the theoretical model typically employed is the so-called "proximity approximation" [13], which is numerically uncontrolled and *ad hoc*. In the literature, there has been substantial discussion regarding the conditions for the validity of this approximation [14].

Further advancement in our understanding of near-field thermal transfer critically relies upon the developments of a mathematically exact and numerically controlled theoretical framework for quantitative analysis of experiments. In this Letter, we present such a framework, which allows one to calculate the electromagnetic heat flux between a sphere (with a spatially homogenous dielectric function $\epsilon_A(\omega)$ and a radius $a$) and a plate (assumed to fill a half space, with a spatially homogenous dielectric function $\epsilon_B(\omega)$), separated by a vacuum gap (with $\epsilon = 1$, and a gap distance $d$). We consider the non-equilibrium situation where the sphere and the plate are maintained at different temperatures, $T_A$ and $T_B$, respectively. A schematic rendering of the geometrical configuration is shown in Fig. 1. The method outputs the exact electromagnetic heat flux, with controlled numerical precision. The method can in principle be applied to other multiple body thermal electromagnetic problems, as well.

In order to calculate the sphere-plate heat flux using CMFED, we introduce, at each point in the sphere, fluctuating electric current sources $j$, which satisfy the fluctuation-dissipation theorem:

$$\langle j_\alpha(\vec{r}_A) j_\beta(\vec{r}_A') \rangle = \frac{1}{\pi} \epsilon_0 \omega \Theta(\omega,T) \text{Im}[\epsilon(\vec{r}_A)] \delta_{\alpha\beta} \delta(\vec{r}_A - \vec{r}_A') \tag{1}$$

Here, $\alpha, \beta = 1\ldots 3$ label the three spatial axes, and $\Theta(\omega,T) = \hbar\omega / \left(\exp(\hbar\omega/k_B T) - 1\right)$. The heat flux $Q$ is then related to the thermal electric field generated by such current sources:

$$Q(T_A,T_B) = \int d\omega \left[\Theta(\omega,T_A) - \Theta(\omega,T_B)\right] \frac{1}{\pi} \frac{\omega^2}{c^2} \operatorname{Im}[\epsilon_1(\omega)] \cdot \operatorname{Re} \int_{V_a} d\vec{r}_A \sum_{\alpha=1}^{3} \int_{\partial V_B} d\hat{n}_B \cdot$$
$$\left[\vec{E}^*(\vec{r}_B,\vec{r}_A,\vec{e}_\alpha,\omega) \times \nabla_B \times \vec{E}(\vec{r}_B,\vec{r}_A,\vec{e}_\beta,\omega)\right] \quad (2)$$

Here, $\vec{E}(\vec{r}_B,\vec{r}_A,\vec{e}_\alpha,\omega)$ is the electric field excited by a point current source with unity strength, located in the sphere at $\vec{r}_A$, with polarization $\vec{e}_\alpha$ and frequency $\omega$. $\nabla_B$ denotes the gradient operator with respect to the $\vec{r}_B$ coordinate. $\vec{E}$ is related to the electric-field dyadic Green function operator $\vec{\vec{G}}$, as

$$\vec{e}_\beta \cdot \vec{E}(\vec{r}_B,\vec{r}_A,\vec{e}_\alpha,\omega) = \langle \vec{r}_B,\vec{e}_\beta | \vec{\vec{G}} | \vec{r}_A,\vec{e}_\alpha \rangle \quad (3)$$

Using the operator notation, the summation over sources in Eq. (2) thus becomes:

$$\operatorname{Re} \int dV_A \sum_{\alpha=1}^{3} \int_{\partial V_B} d\hat{n}_B \cdot \left[\vec{E}^*(\vec{r}_B,\vec{r}_A,\vec{e}_\alpha,\omega) \times \nabla_B \times \vec{E}(\vec{r}_B,\vec{r}_A,\vec{e}_\alpha,\omega)\right]$$
$$= \operatorname{Re} \int_{\partial V_B} d\hat{n}_B \cdot \int dV_A \sum_{\alpha=1}^{3} \left[\langle \vec{r}_B,\vec{e}_\alpha | \vec{\vec{G}}^+ \times \nabla_B \times \vec{\vec{G}} | \vec{r}_A,\vec{e}_\alpha \rangle\right] \quad (4)$$
$$= \operatorname{Re} \sum_{l,m,p} \int_{\partial V_B} d\hat{n}_B \langle l,m,p | \vec{\vec{G}}^+ \times \nabla_B \times \vec{\vec{G}} | l,m,p \rangle$$

Here $|l,m,p\rangle$ is a vector spherical harmonic, where $l$ is the total angular momentum, $m$ is the angular momentum component along the $z$-direction, and $p$ labels the linearly independent solution of the vector wave equation.

As seen in Eq. (4), the key step of the computation is to calculate the flux through the surface of the plate, for a given source distribution inside the sphere as described by $|l_0,m,p_0\rangle$. Noting that $m$ is a conserved quantity in the sphere-plate geometry, we can expand the electric field in the vacuum region, using Einstein's summation convention

$$\vec{E}^{vac}(\vec{r}) = A_\delta \vec{\Omega}_\delta^A(\vec{r}) + B_\gamma \vec{\Omega}_\gamma^B(\vec{r}) \quad (5)$$

.

The $\delta$ and $\gamma$ are multi-indices; $\vec{\Omega}_\delta^A \in \{M_{lm}^{(3)}, N_{lm}^{(3)}\}$ are the outgoing vector spherical harmonics with origin at the center of the sphere, and $\vec{\Omega}_\gamma^B \in \{m_{\lambda m}^{(1)}, n_{\lambda m}^{(1)}\}$ are the vector cylindrical harmonics reflected from the plate, with $\lambda$ as the cylindrical wavevector along the radial direction. In the following, $\vec{I}_\gamma^B \in \{m_{\lambda m}^{(1)}, n_{\lambda m}^{(1)}\}$ are the incident vector cylindrical harmonic waves traveling toward the plate, and $\vec{I}_\delta^A \in \{M_{lm}^{(1)}, N_{lm}^{(1)}\}$ are the vector spherical harmonic waves traveling from the plate toward the sphere.

It is useful to introduce the operators $\Lambda^{A \leftarrow B}$ and $\Lambda^{B \leftarrow A}$, which relate the vector spherical and cylindrical harmonics. Closed form integral expressions exist in the literature [15]. In Eq. (5), we may insert the relation $A_\delta \vec{\Omega}_\delta^A = \Lambda_{\gamma\delta}^{B \leftarrow A} A_\delta \vec{I}_\gamma^B$ to arrive at:

$$B_\gamma = R_\gamma^B \Lambda_{\gamma\delta}^{B \leftarrow A} A_\delta \qquad (6)$$

where $R_\gamma^B$ are the reflection coefficients for the cylindrical waves at the plate surface. Similarly, we may insert the relation $B_\gamma \vec{\Omega}_\gamma^B = \Lambda_{\delta\gamma}^{A \leftarrow B} B_\gamma \vec{I}_\delta^A$ into Eq. (5) to obtain:

$$A_\delta = A_\delta^0 + R_\delta^A \Lambda_{\delta\gamma}^{A \leftarrow B} B_\gamma \qquad (7)$$

where $R_\delta^A$ are reflection coefficients for the spherical waves at the boundary of the sphere. The $A_\delta^0$ are coefficients of the outgoing spherical harmonics due to the sources inside the sphere. These $A_\delta^0$ can be calculated analytically from the dyadic Green's function of an isolated sphere, which are well known.

Combining Eqs. (6) and (7), we have:

$$(I_{\delta\delta'} - R_\delta^A \Lambda_{\delta\gamma}^{A \leftarrow B} R_\gamma^B \Lambda_{\gamma\delta'}^{B \leftarrow A}) A_{\delta'} = A_\delta^0 \qquad (8)$$

The solution to Eq. (8) determines the electric field in the vacuum region via Eq. (5), and hence the flux into the plate for a given source distribution described by $|l_0, m, p_0\rangle$. The sum of the incoherent erngy flux contributions from all possible source configurations yields the total heat flux via Eqs. (2) and (4).

In the numerical implementation of Eq. (8), the summation over the $\delta$, $\gamma$ indices are carried out for $l$ up to some $l_{max}$, and for $m$ up to some $m_{max}$. To achieve ~1% numerical convergence for a system with $a = 20\mu m$, $d = 100 nm$, which is typical of the experiments [9-10], requires $l_{max} \sim 700$. Hence, Eq. (8) represents a modestly sized linear system, even in such an extreme geometry with $a/d = 200$. More detailed information regarding the convergence of the algorithm can be found in the supplementary information.

We can directly compare our results with the experimental heat flux data from [10], which was measured from a silica sphere-plate system at $T_1 = 321K$, $T_2 = 300K$, with $a = 20\mu m$ and $20 nm < d < 3\mu m$. For reference, the thermal wavelength $\lambda_T \equiv \hbar c / k_B T$ as well as the electromagnetic surface resonance wavelengths for silica surfaces are on the order of $10\mu m$. Our results are consistent with the data in the same sense that the standard proximity approximation is [10]. That is, the uncertainties in the experimental calibrations of parameters are large enough to allow for very good fits to both theories. Details of this comparison can be found in the supplemental information.

It is straightforward to explore a wider range of the $a - d$ parameter space using our method. In Fig. 2 we plot the exact heat flux $Q$ as a function of sphere radius $a$, for various fixed gap distances $d$. We compare our exact results with three commonly used approximations in order to provide a unifying picture across different regimes. Such a comparison generates considerable insights into near-field electromagnetic heat transfer in general.

*Far-field approximation.* In Fig. 2, we normalize the values for the heat flux to the far-field limit, $Q_{ff}$, which we define as the heat flux in the absence of interference due to secondary reflection from the sphere. In general, the exact $Q$ deviates from $Q_{ff}$ only when $d$ is sufficiently small. The near-field contribution is significant only when $d \lesssim 1 \mu m$ (see Fig. 2, $d = 100 nm$, $1000 nm$).

Perhaps somewhat surprisingly, our results reveal that for every gap size $d$, $Q$ always approaches $Q_{ff}$, in the limit of large $a$. This is true even when the gap size is as small as $d = 100 nm$, when one expects a substantial near-field contribution. This behavior arises from the specific geometric aspects of the sphere-plate configuration. In particular, for sufficiently small gaps, the near-field contribution scales as $a/d$, while the far-field contribution scales as $a^2$. Thus the far field contribution dominates when the radius of the sphere is large. This observation is relevant for experimental design: In order to achieve substantial enhancement of heat flux above the far-field limit, it is important to use small, micron (or sub-micron) scale spheres, as can be inferred from Fig. 2.

*Dipole approximation.* In Fig. 2, $Q_{dipole}$ is the result from the dipole approximation, which treats the sphere as a single point dipole emitter [16]. The dipole approximation is equivalent to fixing $m_{max} = 1$ in our method. Referring to Fig. 2, we find that for any given $d$, $Q_{dipole}$ asymptotically approaches $Q$ when $a \ll d$, and that the dipole approximation is actually quite good over a substantial range of radii. For example, when $d = 1 \mu m$, the dipole approximation results deviate from the exact results by less than 10%, even for $a$ as large as $a \approx d = 1 \mu m$. The approximation deviates from the exact value only when $a > d$. Since the enhancement over the far-field limit is most prominent when the sphere is small, our results here indicate that the dipole approximation is useful in regimes where enhancement over the far field limit is substantial.

*Proximity approximation.* In Fig. 2, $Q_{prox}$ is the result from the proximity approximation [10,13], which assumes that all heat transfer occurs pointwise, between closest points on the sphere and plate. Mathematically:

$$Q_{prox}(a,d) = \int_0^a dr 2\pi r Q_{pp}(d + a - \sqrt{a^2 - r^2})  \qquad (9)$$

where $Q_{pp}(h)$ is the heat flux surface density for two plates separated by a distance $h$, which is known analytically [1].

Referring again to Fig. 2, we observe that for small radii ($a \ll d$, a regime very relevant to enhancement over the far field limit), the proximity approximation consistently overestimates the heat flux for two related reasons. First, it ignores the effect that the spherical curvature has on scattering. Second, it does not distinguish between propagating and evanescent waves. As recognized in the literature, the proximity approximation does provide a useful approximation when $a > d$. However, when $a \gg d$, the proximity approximation does not approach the correct far-field limit; it underestimates the far-field contribution due to emission by the part of the sphere far from the plate. Since it fails in both extremes of radius, it is difficult to interpret the physical meaning of the proximity approximation.

As another illustration, we plot in Fig. 3 the heat flux as a function of $d$ for fixed $a = 1\mu m$. We notice that the proximity and dipole approximations are valid when $d < a$ and $d \gg a$, respectively. The cross-over between the two regimes occurs near $d \approx 300 nm$; none of the standard approximations apply in this case.

To summarize, in Fig. 4 we present the "phase diagram" of the $(a,d)$ parameter space in terms of the validity of various approximations. Our calculations provide a unified view for various approximations that have been previously developed to treat this important geometry. While we have focused on the sphere-plate geometry in this letter, the method we have outlined can be straightforwardly generalized to other multiple-body geometries, provided the scattering matrix for each individual body is known. The method is therefore of general applicability for understanding a variety of near-field electromagnetic effects, including Casimir effects, optical forces, and vacuum friction.

This work is supported in part by an AFOSR-MURI program, (Grant No. Grant No. FA9550-08-1-0407), and by the U. S. Department of Energy, Office of Basic Energy Sciences, Division of Material Sciences and Engineering. The computations were carried out through the support of the NSF-Teragrid program.

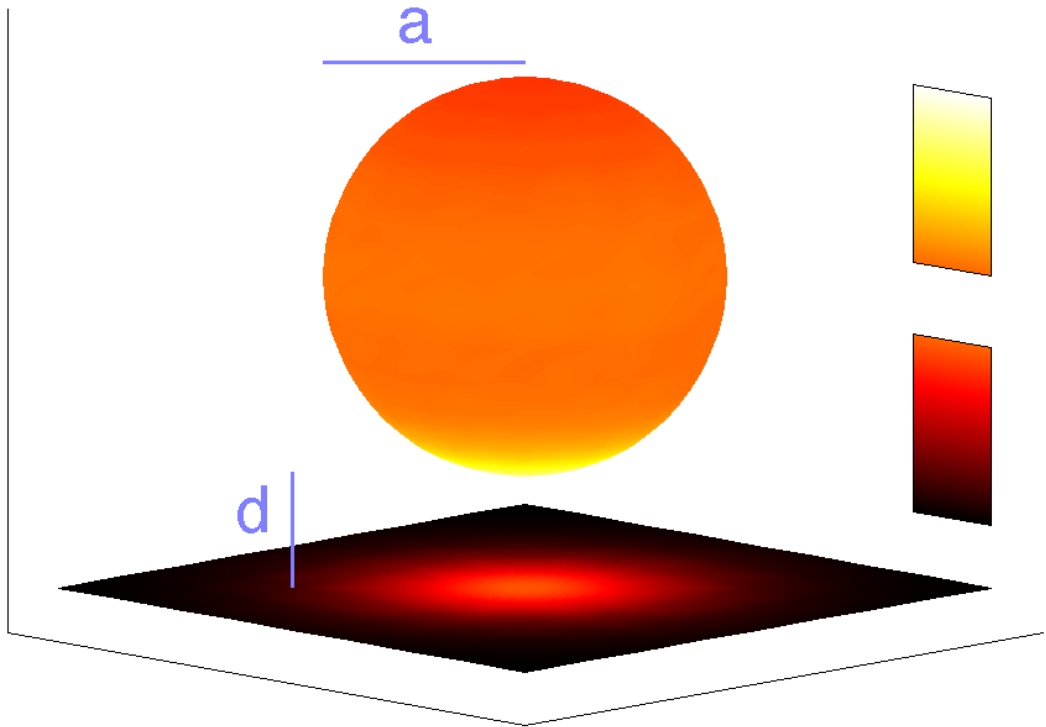

Figure 1. A schematic of the geometry considered: a dielectric sphere of radius $a$, and a dielectric plate separated by a distance $d$ through a vacuum gap. The color corresponds to electric field intensity near the surface of the sphere and the plate, using the parameters $a = 1{,}000\,nm$, $d = 800\,nm$ as a representative example. The field near the sphere is much greater than that near the plate, so in order to resolve the variation in intensity, the scales (see the colorbars on the right) are not the same. In all the examples in this letter, the sphere and plate are both amorphous $SiO_2$, the sphere is at temperature $T_1 = 321K$, and the plate is at $T_2 = 300K$.

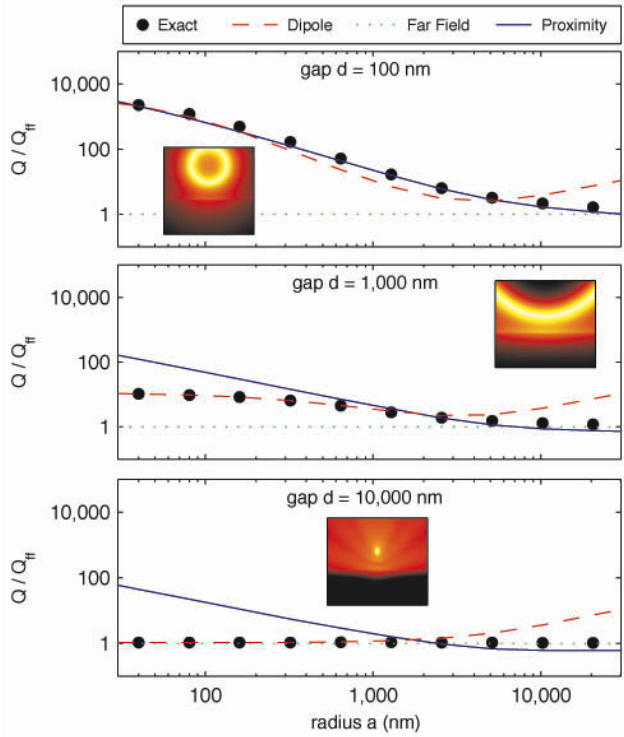

Figure 2. Heat flux as a function of $a$ for fixed $d = 100 nm, 1 \mu m, 10 \mu m$. Insets are calculated scattered electric field intensities (located in the figure at the appropriate value of $a,d$). Each curve is normalized by the far-field heat flux $Q_{ff}$, which is by definition independent of $d$.

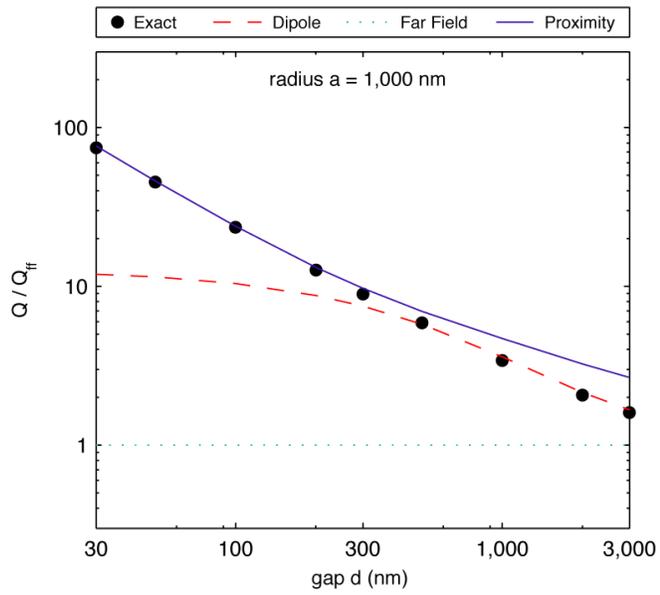

Figure 3. Exact heat flux compared with dipole and proximity approximations, as a function of $d$, with fixed $a = 1\mu m$.

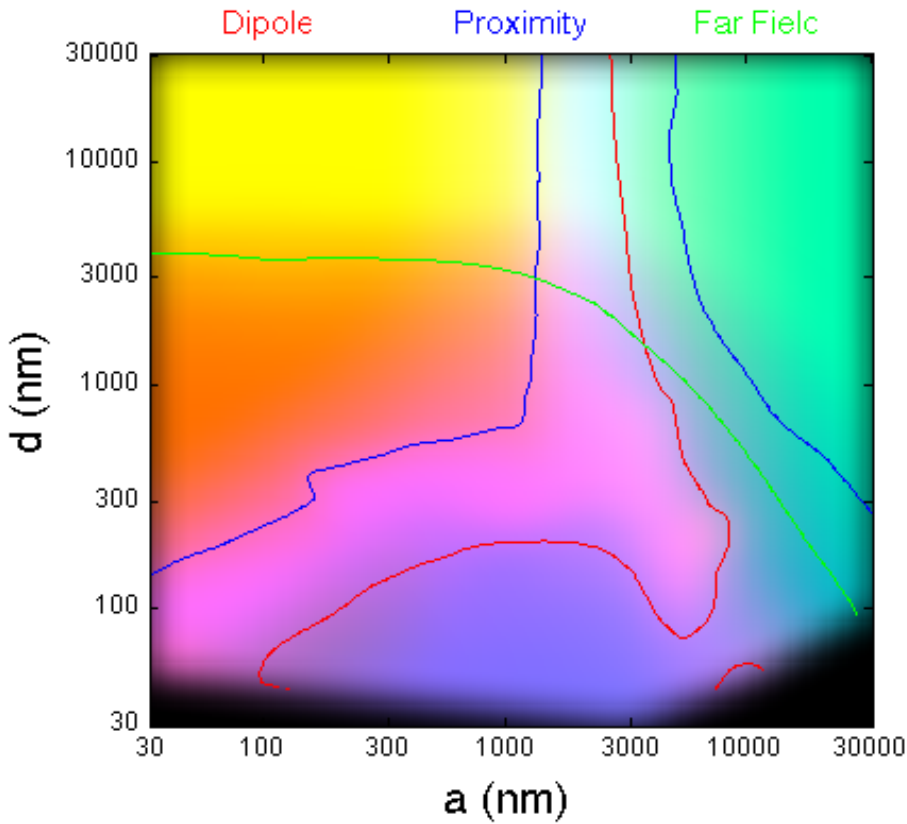

Figure 4. An RGB color-coded $(a,d)$ phase-space diagram of the relative accuracy of three important approximations to the sphere-plate heat flux. The value of the red, green, or blue color component of each pixel corresponds to the error as $\log(Q_{exact}/|Q_{approximate} - Q_{exact}|)$ where $Q_{approximate}$ corresponds to the dipole, far-field, or proximity approximations, respectively. The minimum error (corresponding to the maximum value of the color component) is 10%. The maximum error (corresponding to color component) is 450%. Thus, for example, the green region corresponds to a regime where the far-field approximation is valid. Colors are additive in the usual sense, so, for example, the yellow region corresponds to a regime where the dipole (red) and far-field (green) approximations are both valid. The color-coded contour lines denote 30% error; the arrows point toward the region of validity. The image is rendered from a smoothed interpolation of 147 data points.

# Exact microscopic theory of electromagnetic heat transfer between a dielectric sphere and plate

Clayton Otey[1] and Shanhui Fan[2]

[1] Department of Applied Physics and [2]Department of Electrical Engineering, Stanford University, Stanford, California, 94305 U.S.A.

**Convergence**

Here we give a brief physically motivated convergence analysis. In Supplementary Figure 1, we break up the heat flux (calculated using the geometrical parameters from [10]) into contributions $Q(m)$ and $Q(l)$,

$$Q(m) = \sum_{l} Q(m,l)$$
$$Q(l) = \sum_{m} Q(m,l) \tag{S1}$$

where $Q(m,l)$ is the contribution to $Q$ for a given $m$, $l$ in Eq. (4). The basic expectation is that reasonable convergence should be achieved when $l_{max}$ is in some sense large enough to resolve the relevant scales. In general, the far-field contributions scale as $(a\omega/c)$, and the near-field contributions scale as $a/d$ [12]. In our implementation we set $l_{max} = \kappa_0 + \kappa_1(a\omega/c + a/d)$ with $\kappa_0 = 8$, $\kappa_1 = 2.5$, and we find behavior similar to that shown in the supplementary figure in all of the data presented in the main document. If we instead choose $\kappa_1 = 2.0$ or $\kappa_1 = 3.0$, the results change by less than 2%. In general, the error decays rapidly as a function of $\kappa_1$ for $\kappa_1 \geq 2.0$.

When carrying out the frequency integration of Eq. (1) in the main text, we sample the frequency space by explicitly choosing frequencies that resolve the relevant surface resonances. In all calculations presented in this letter, we used 47 frequencies.

**Data fit to the experiment in Ref. [10]**

In the experiment described in [10], a silica sphere is attached to a cantilever. The deformation $\Delta(d)$ of the cantilever is then measured as the sphere is brought to be in the vicinity of a silica plate, where $d$ is the sphere-plate vacuum gap. The heat flux between the sphere and plate is related to such deformation through:

$$Q(d) - Q(d \to \infty) = H \cdot \Delta(d - d_0) \tag{S2}$$

$H$ and $d_0$ are experimental calibration parameters. Measurements of similar cantilevers suggested $H = 2.30 \pm 0.69 nW/nm$ [10]. The characteristic surface roughness was $40 nm$, and hence there is uncertainty in $d_0$ as well. In comparing theoretical calculations of $Q(d)$ with such experiments, we note that the experiments do not directly measure $Q(d \to \infty)$.

Based on their experimental $\Delta(d)$ data [10], Greffet et al. reported an excellent fit of the proximity approximation to $Q(d)$ in Eq. (S2), using the parameters $Q(d \to \infty) = 5.45 nW$, $H = 2.162 nW/nm$, $d_0 = 31.8 nm$. Our exact results of $Q(d)$ do not agree with the proximity approximation, as seen in Supplementary Fig. 2a. However, as seen in Supplementary Fig. 2b, our exact results for $Q(d)$ can be fit to the experimental $\Delta(d)$ data as well, with the choice of parameters: $Q(d \to \infty) = 9.74 nW$, $H = 2.134 nW/nm$, $d_0 = 32.3 nm$. We note that the differences between these two sets of parameters are well within experimental uncertainties. The fact that the same experimental data can be fit to two very different theoretical models should motivate a more thorough assessment of uncertainties in precision nano-scale sphere-plate electromagnetic heat transfer experiments.

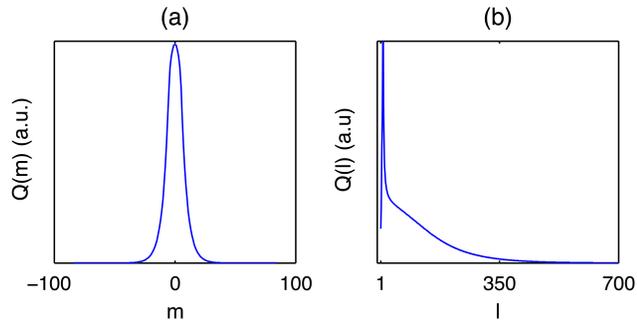

**Supplementary Figure 1**. Typical contribution to the spectral energy flux as a function of $m$ and $l$ ( $a = 20\mu m$, $d = 100 nm$, $\hbar\omega = .0605 eV$ (corresponding to a surface phonon polariton resonance), $\epsilon_1 = \epsilon_2$ corresponds to amorphous $SiO_2$ ).

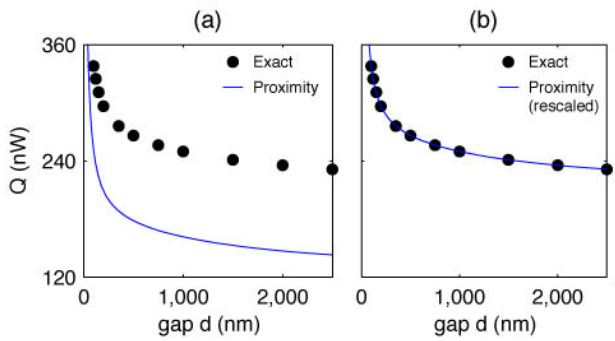

**Supplementary Figure 2.** a) Our results do not agree with the proximity approximation. b) The proximity approximation (and the experimental data from [10]) can be scaled to match our results.